\documentclass{aa}  
\usepackage{txfonts}

\usepackage[toc,page]{appendix}

\usepackage[table]{xcolor}
\usepackage[utf8]{inputenc} 
\usepackage[T1]{fontenc}    
\usepackage{hyperref}       
\usepackage{url}            
\usepackage{booktabs}       
\usepackage{amsfonts}       
\usepackage{nicefrac}       
\usepackage{microtype}      
\usepackage{footmisc}
\usepackage{graphicx}
\usepackage{natbib}

\usepackage{tabu}
\usepackage{makecell}
\usepackage{listings}
\usepackage{amsmath}
\usepackage{color}
\usepackage{multirow}
\usepackage{xcolor}

\usepackage{multiobjective}

\newcommand{\quotes}[1]{``#1''}

\usepackage{placeins}
\usepackage{cleveref}

\usepackage{ulem}

\makeatletter
\def\@fnsymbol#1{\ensuremath{\ifcase#1\or \dagger\or \ddagger\or
   \mathsection\or \mathparagraph\or \|\or **\or \dagger\dagger
   \or \ddagger\ddagger \else\@ctrerr\fi}}
\makeatother

\begin{document} 
   \title{A baseline on the relation between chemical patterns and birth stellar cluster}
   \author{T. Signor\inst{1,2}\thanks{\href{theosamuele.signor@mail.udp.cl}{\tt theosamuele.signor@mail.udp.cl}}\and
   P. Jofré\inst{1,3}\and
   L. Martí\inst{2}\and
   N. Sánchez-Pi\inst{2}
          }
\institute{
Instituto de Estudios Astrofísicos, Facultad de Ingeniería y Ciencias, Universidad Diego Portales, Av. Ejercito 441, Santiago, Chile
\and
Inria Chile Research Center, Av. Apoquindo 2827, piso 12, Las Condes, Santiago, Chile
\and
Millenium Nucleus ERIS}

            
\date{Received March 11, 2024; accepted May 22, 2024}

 
  \abstract
   {The chemical composition of a star's atmosphere reflects the chemical composition of its birth environment. Therefore, it should be feasible to recognize stars born together that have scattered throughout the galaxy, solely based on their chemistry. This concept, known as \quotes{strong chemical tagging}, is a major objective of spectroscopic studies, but has yet to yield the anticipated results.}
   {We assess the existence and the robustness of the relation between chemical abundances and birth place using known member stars of open clusters.}
   {We followed a supervised machine learning approach, using chemical abundances obtained from APOGEE DR17, observed open clusters as labels and different data preprocessing techniques.}
   {We found that open clusters can be recovered with any classifier and on data whose features are not carefully selected. In the sample with no field stars, we obtain an average accuracy of $75.2\%$ and we find that the prediction accuracy depends mostly on the uncertainties of the chemical abundances. 
   When field stars outnumber the cluster members, the performance degrades.
      }
   {
   Our results show the difficulty of recovering birth clusters using chemistry alone, even in a supervised scenario. This clearly challenges the feasibility of strong chemical tagging. Nevertheless, including information about ages could potentially enhance the possibility of recovering birth clusters.}
   \maketitle
\section{Introduction}\label{sec:intro}
Most of the star formation in the Milky Way is believed to take place in chemically homogeneous collapsing molecular clouds \citep{ladalada2003embedded}, forming stellar aggregates such as open clusters or unbound associations. These birth aggregates are characterized by having similar chemical compositions, ages, and motions through space \citep[e.g.][and references therein]{Soubiran18, GaiaESO}. 

However, as most stars are born in weakly-bound association and thus rapidly dispersed and separated from their siblings due to gravitational interaction \citep{Krumholz2019starclusters}, the motions of the members of these groups quickly lose all the information about the birth place.
The chemical composition of the atmosphere of low to intermediate mass stars, on the other hand, remains largely constant throughout their life. This means that, if the cloud is actually chemically homogeneous \citep{feng2014early, BlandHawthorn2010long}, it should be possible to use their chemical patterns as a fingerprint of their birth environments and birth stellar associations, even when these are distrupted \citep{Freeman2002newgalaxy}. The task of uniquely and confidently assigning individual stars to specific birthplaces is referred to as ``strong chemical tagging''. 

Some of these birth stellar associations are massive enough to remain gravitationally bound from millions to billions of years and these are the open clusters we observe today. Observational studies have been a crucial part of testing the feasibility of strong chemical tagging. Surveys like APOGEE (\citealt{APOGEE}), LAMOST (\citealt{LAMOST}), GALAH (\citealt{GALAH}), RAVE (\citealt{RAVE}), and Gaia-ESO (\citealt{GaiaESO}) have aimed to gather data on the chemical composition of stars in these open clusters. These datasets allow us to test if there is a relation between chemical patterns and birth association and if this relation can be estimated from observations.

One major challenge of chemical tagging is the complexity of the chemical evolution of the galaxy. While the chemical composition of a star depends on its birthplace, the chemical enrichment history of that birthplace has a further dependency on the chemical evolution of the region in which it formed \citep{Magrini23,Vitali24}.  This evolution might be further affected by inhomogeneous mixing of gases, accretion of smaller galaxies and stellar migration \citep{Bird12, Johnson21}. This means that stars formed in different regions of the galaxy may have similar chemical compositions \citep{Edvardsson93}, making it difficult to confidently assign a specific birthplace to an individual star.

The accuracy of the observational data poses another significant challenge for chemical tagging \citep{BlancoCuaresma2015testing, Bovy16}. High-resolution spectroscopy is needed to measure the detailed chemical fingerprints of individual stars, but the precision and accuracy of these measurements are still limited \citep[see e.g.][for a review]{Jofre19}. This means that the chemical fingerprints of some stars may not be distinct enough to disentangle different birth places \citep{Ness18, Manea24}.

Previous attempts at addressing strong chemical tagging have been focused in performing a machine learning clustering analysis on some chemical abundance space to evaluate if observed open clusters are recovered. Overall, they are about applying clustering analysis techniques in order to perform unsupervised classification, as they make use of known cluster members to evaluate the recovery performance. Unsupervised classification involves grouping (clustering) and distributing stars based solely on their similarity, computed through a distance metric, in some chemical abundances space. Then a class, or label, is assigned to each of the groups found. Essentially, these methods aim to identify statistical groups of stars sharing similarities, with an expectation that these would somewhat correspond to actual clusters.

For example, one of the first attempts to perform strong chemical tagging following \cite{BlancoCuaresma2015testing} was presented in \cite{Price-Jones2020}, where a sample of more than $180\,000$ stars belonging to all the Galactic components was used. The reported inability of recovering any of the known open clusters present in their data set stresses the difficulty of the task. This is especially true when noisy data significantly blur and wash out the chemical patterns we rely on \citep[see e.g. discussions in][]{Bovy16}. 

To address this last limitation, the studies from \cite{Casamiquela_2021} and \cite{Spina} used precise chemical abundance estimates for member stars in open clusters to group them using the density-based clustering algorithms HDBSCAN \citep{HDBSCAN} and OPTICS \citep{Ankerst1999optics}. 
Despite using higher-quality data and sophisticated clustering methods, they were not able to recover open clusters with more than 50\% completeness and homogeneity. Indeed, most of the groups found were statistical groups containing stars belonging to different open clusters. These recent works thus support previous conclusions of the impossibility of strong chemical tagging.

How the various groups found with clustering algorithms are computed depends on the adopted algorithm and other parameters defined beforehand by the user, such as the threshold for similarity \citep{Casamiquela_2021}. The real classes are never provided to the clustering algorithm. Furthermore, the clustering very strongly depends on the feature space and the samples provided to the algorithm. This has a larger impact on the result than the specific algorithm being used \citep[e.g.][]{Manea24}.

In this work, differently from the ones mentioned above, we aim to give a step prior to clustering, which is to classify to evaluate if a grouping of abundance data of open clusters is possible.  
We reframe the task as a supervised classification problem, as we are making use of the labels of class (the truth labels $y$) to learn the mapping $y=f(\vec{x})$ from the chemical features $\vec{x}$ to these labels $y$. Once this function is estimated from some training data, we can use it to to reconstruct it or making prediction on unknown data and thus assess its quality.

The whole idea of applying this kind of analysis is to shift the question from ``can we group stars together in chemical space to recover birth clusters?'' to ``can we connect chemistry to stars' birth place?''.
Supervised classification, with its greater accuracy compared to its unsupervised counterpart, will set an upper limit to our hopes to perform strong chemical tagging with the currently available data sets.
Furthermore, finding a supervised learning link might open the door to eventually finding a representation where open clusters can be found unsupervisingly.

Summarizing, the goal of this work is not to attempt to recover disrupted clusters, but to evaluate the feasibility of this task from an empirical perspective, by trying to answer in a systematic way the following questions:
\begin{enumerate}
    \item Is there actually a link between chemical abundances and birth open clusters?
    \item What are the most important factors of variation in the recovery of open clusters?
    \item Which are the most important chemical elements?
    \item Is it possible to find a better representation for these chemical abundances in a physics-agnostic way?
\end{enumerate}

The structure of the paper is then as follows. In Sect.~\ref{sec:data} the data used are introduced. In Sect. \ref{sec:methods} the main methods to achieve our goals are described, followed by the results obtained in Sect.~\ref{sec:results}. Finally, Sect.~\ref{sec:discussion} concludes this paper summarizing the main results and the perspectives for future work.

\section{Data}\label{sec:data}
We made use of the seventeenth data release of the APOGEE survey (\citealt{DR17}). The catalog includes data for a total of about $657\,000$ stars, both in the field and in clusters. The stars are observed with the APOGEE spectrograph, which is an infrared instrument with a resolving power of about 20,000, allowing the determination of radial velocities (RV) \citep{Nidever15}, atmospheric parameters and abundances of about 20 different chemical species \citep{APSCAP}. 

To assess whether each single star is member of an open cluster, we use of membership probabilities from the Open Cluster Chemical Analysis and Mapping (OCCAM) catalog (\citealt{OCCAM1, OCCAM2, Myers2022open}). This is obtained by combining the provided APOGEE RVs and metallicities ([Fe/H]) with proper motion observations (PM) from the Gaia Data Release 3 \citep{Vallenari2023gaiadr3} to establish membership probabilities for APOGEE stars in open cluster fields. 
The DR17 version of this catalog contains a total of 153 open clusters and some of them are flagged as ‘‘high quality’’ based on the appearance of their color-magnitude diagram. 
We consider members of open clusters all the stars with $p_{\rm PM}>0.6,\,p_{\rm RV}>0.6,\,p_{\rm [FE/H]}>0.3$ or with $p_{\rm PM}\times p_{\rm RV}>0.25,\,p_{\rm CG}>0.85$. Here $p_{\rm PM}$, $p_{\rm RV}$, and $p_{\rm [Fe/H]}$ represent the membership probabilities based on proper motion, radial velocity, and metallicity, respectively. $p_{\rm CG}$ indicates the membership probability from \citealt{CG2018}. We also discard clusters with less than eight observed members. These thresholds and cuts are arbitrary, involving a trade-off between maintaining an adequate sample size and ensuring quality in membership assessments.

For what concerns the chemical abundances, our input features, we adopt those obtained from {\tt AstroNN} (\citealt{AstroNN}). This data analysis pipeline provides abundance determinations (along with their uncertainties $\sigma$) of all the following elements: C, Ci\footnote{Ci represents the abundance of atomic carbon only, while C represents the overall carbon abundance, which includes both atomic and molecular carbon.}, N, O, Na, Mg, Al, Si, P, S, K, Ca, Ti, Ti II, V, Cr, Mn, Fe, Co and Ni, expressed as the logarithm of the ratio of a star's element abundance with the hydrogen abundances and compared to that of the Sun (indicated with $\odot$):
\begin{align}
    \left[ \frac{X}{H}\right] = \log_{10} (\frac{N_X}{N_H})-\log_{10} (\frac{N_X}{N_H})_{\odot}\, ,
    \label{eq:X_H}
\end{align}
where $N_X$ and $N_H$ are the number of element $X$ and hydrogen atoms per unit of volume, respectively. 
While some of these elements, like carbon and nitrogen, can be modified by stellar evolution \citep{masseron2015carbon,salaris2015post}, we remind how our supervised classification approach, contrarily to its unsupervised counterpart, should be able to mitigate this concern by learning to assign appropriate weights to each feature during training. 
We show deviations from the cluster mean for each element as a function of effective temperature in Fig. \ref{fig:trendsTeff}, finding that variations in this parameter do not introduce systematic trends within the chemical abundances of cluster members.
Notably, carbon and nitrogen abundances do not exhibit significantly larger deviations than other elements. This could be due, in part, to a selection bias favoring unevolved stars within our cluster member sample, where mixing has not yet substantially altered the abundance of these elements.

Our motivation to use this dataset is that it has higher precision compared to the standard data releases of APOGEE because the results are obtained after training a neural network on a sample of well determined abundances from ASPCAP \citep{APSCAP}. This further ensures consistency between the membership probabilities and the chemical abundances used in this work, since only precision is improved, not accuracy. 

We discarded phosphorous as a feature because of its errors, and stars with ${ \rm [Fe/H]} <-0.6\, \mathrm{dex}$ and mean chemical abundances uncertainties $>0.6\, \mathrm{dex}$.  
After all these selections and quality cuts we count $m=775$ stars in 27 clusters. In Table \ref{table:counts} we show the resulting number of members per cluster. 

\begin{table*}
\caption{Members counts, average metallicity, galactic radius and age for clusters in our selection.}
\label{table:counts}
\centering
\begin{tabular}{lccccc}    
\toprule          
\textbf{Cluster} & \textbf{[Fe/H]} & \boldmath{$R_{\rm GC}$}\textbf{ [kpc]}&\textbf{log(Age/yr)}&\textbf{Members} & \textbf{Frequency}\\
\midrule
Berkeley 17 & $-0.17\pm 0.03$ & 10.62 & $10\pm 0.14$ & 8 & 1 \\ 
NGC 1817 & $-0.18\pm 0.03$ & 9.46 & $ 9.09\pm 0.02$ & 8 & 1 \\
NGC 6705 & $0.12\pm 0.02$ & 6.23 & $ 8.47\pm 0.09$ & 8 & 1 \\ 
Trumpler 5 & $-0.46\pm 0.02$ & 10.42 & $9.54 \pm 0.02$ & 8 & 1 \\ 
Melotte 22 & $-0.05\pm 0.12$ & 8.12 & $ 8.12\pm 0.11$ & 9 & 1.1 \\ 
ASCC 16* & $-0.13\pm 0.09$ & 8.33 & $ 7.09\pm 0.06$ & 11 & 1.4 \\
NGC 2243 & $-0.49\pm 0.03$ & 9.96 & $9.54 \pm 0.04$ & 11 & 1.4 \\ 
Collinder 69* & $-0.12\pm 0.06$ & 8.56 & $ 6.95\pm 0.05$ & 12 &1.5 \\
IC 348* & $-0.11\pm 0.12$ & 8.27 & $ 6.78\pm 0.07$ & 13 & 1.6 \\ 
ESO 211 03 & $-0.19\pm 0.02$ & 8.99 & $ 8.89\pm 0.04$ & 14 & 1.8 \\
ASCC 19* & $-0.12\pm 0.12$ & 8.33 & $ 7.14\pm 0.03$ & 15 & 1.9 \\
IC 166 & $-0.11\pm 0.03$ & 10.69 & $9.01 \pm 0.03$ & 16 & 2.0 \\ 
Ruprecht 147 & $0.09\pm 0.06$ & 7.73 & $ 9.45\pm 0.02$ & 18 & 2.3 \\
NGC 2420 & $-0.20\pm 0.03$ & 9.86 & $ 9.35\pm 0.02$ & 20 & 2.6 \\
NGC 1245 & $-0.12\pm 0.03$ & 10.27 & $ 9.08\pm 0.02$ & 21 & 2.7 \\
NGC 2204 & $-0.30\pm 0.03$ & 10.46 & $ 9.32\pm 0.03$ & 21 & 2.7 \\
Trumpler 20 & $0.09\pm 0.02$ & 6.95 & $ 9.18\pm 0.02$ & 21 & 2.7 \\
Kharchenko 1 & $-0.04\pm 0.04$ & 8.81 & $ 8.76\pm 0.09$ & 24 & 3.1 \\
Berkeley 18 & $-0.38\pm 0.03$ & 11.98 & $ 9.61\pm 0.07$ & 26 & 3.3 \\
BH 56* & $-0.09\pm 0.10$ & 8.13 & $ 7.16\pm 0.08$ & 28 & 3.6 \\
NGC 6819 & $0.03\pm 0.03$ & 7.69 & $ 9.42\pm 0.03$ & 37 & 4.8 \\
NGC 752 & $-0.07\pm 0.05$ & 8.29 & $ 9.18\pm 0.03$ & 37 & 4.8 \\
NGC 188 & $0.06\pm 0.03$ & 8.91 & $ 9.79\pm 0.04$ & 42 & 5.4 \\
NGC 2158 & $-0.25\pm 0.03$ & 11.22 & $ 9.22\pm 0.12$ & 53 & 6.8 \\
NGC 6791 & $0.29\pm 0.04$ & 7.54 & $ 9.86\pm 0.04$ & 65 & 8.4 \\
NGC 7789 & $-0.03\pm 0.06$ & 8.89 & $ 9.21\pm 0.01$ & 68 & 8.8 \\
NGC 2682 & $-0.00\pm 0.03$ & 8.56 & $ 9.58\pm 0.03$ & 161 & 20.8 \\
\bottomrule
\end{tabular}
\tablefoot{The metallicity and its relative uncertainty are computed as a weighted average and weighted standard deviation of the metallicities of the individual members, respectively, where the weights are inversely proportional to the square of the uncertainties.
\textit{Frequency} indicates the percentage of stars in a particular cluster compared to the total data size. $R_{\rm GC}$ indicates the distance from the galactic center, as obtained using parallaxes from \cite{GaiaDR2}. Ages taken from \cite{dias2021updated}. Ages for Berkeley 18 and NGC 2158 from \cite{Netopil2016metallicity}. (*) clusters flagged as non high-quality \citep{OCCAM1}.}
\end{table*}

\begin{figure*}
    \centering
    \includegraphics[width=\linewidth]{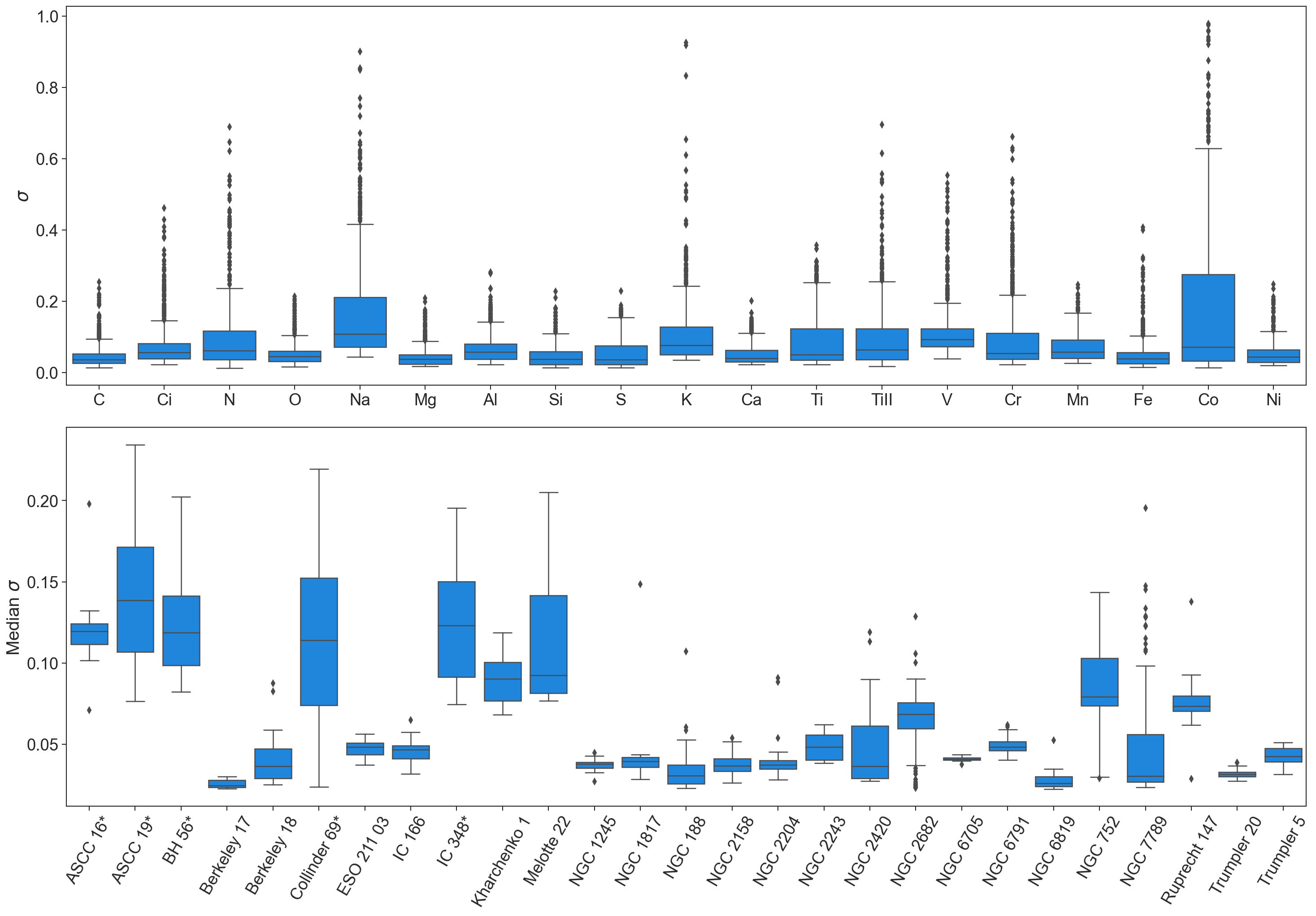} 
    \caption{Distributions of uncertainties. (\textbf{Top}) Distribution of the abundance uncertainties of the different elements for individual stars.  (\textbf{Bottom}) Distribution over cluster members of the median (over chemical abundances) uncertainties.
    In both panels, each box plot is relative to the group indicated in the $x$-axis. Each blue box and contains the middle 50\% of the data and the horizontal line within the box denotes the median value. Outliers are indicated as individual points.}
    \label{fig:ab_uncertainties}
\end{figure*}

The distribution of the abundance uncertainties of the different elements is shown in the top panel of Fig. \ref{fig:ab_uncertainties}, with horizontal lines indicating the median value. 
In the bottom panel of Figure \ref{fig:ab_uncertainties}, we plot the distribution of abundance uncertainties (i.e. the median uncertainty across elements) for each cluster. Members of high-quality clusters show mostly uniform uncertainties, often with a dispersion less than $\approx 0.05\, \mathrm{dex}$. In contrast, clusters not flagged as high-quality tend to display greater variability, as exemplified by Collinder 69, which has a dispersion around $\approx0.12\, \mathrm{dex}$). 

Since our purpose is to classify open clusters using chemistry only, we refer to a vector of chemical abundances as $\bf{x}$, and the cluster name as the label $y$. For now on we will  thus see our data as vectors and labels to study the various classifier methods.
\section{Methods}\label{sec:methods}
The problem of finding a mapping from chemical abundances to birth stellar cluster is recasted as a classification problem: the question is if it is possible to represent the open cluster label $y$ as a function $f$ of the chemical abundances  $\bf{x}$, i.e.
\begin{align}
    y_i\approx f(\bf{x_i})\,,
\end{align}
where $i=1,\ldots,m$ runs over samples. In this section we present the main methods employed in this work, which include the performance metrics and the evaluation methods, the classifiers and the choices of the feature space.

\subsection{Performance metrics}
Four metrics are used to evaluate the cluster classification performance. In the following definitions, $\alpha$ indicates a specific true label (i.e. a specific open cluster),  $\text{True Positives}(\alpha)$ is the number of samples that are correctly classified with label $\alpha$, $\text{False Positives}(\alpha)$ is the number of samples that are incorrectly classified as positive for label $\alpha$, $\text{False Negatives}(\alpha)$ is the number of instances that are incorrectly classified as negative for label $\alpha$.
\begin{itemize}
    \item the \emph{accuracy score} is the ratio between the number of correct predictions and the total number of predictions
    \begin{align}
        \text{Accuracy} =
        \frac{\text{Number of correct predictions}}{\text{Number of predictions}}
    \end{align}
    \item the \emph{$\text{recall}(\alpha)$} indicates the ratio of clusters members of cluster $\alpha$ correctly assigned to cluster $\alpha$. 
    \begin{align}
        \text{recall}(\alpha)=
        \frac{\text{True Positives}(\alpha)}{\text{True Positives}(\alpha) + \text{False Negatives}(\alpha)}
    \end{align}
    A high recall($\alpha$) indicates it is possible to identify most of (or all) the members of the cluster $\alpha$ in the data, while a low recall($\alpha$) means that it is not possible to identify siblings born in cluster $\alpha$. 
    \item the \emph{$\text{precision}(\alpha)$} indicates the fraction of cluster members assigned to cluster $\alpha$ with respect to the total number of predictions in cluster $\alpha$.
    \begin{align}
        \text{precision}(\alpha)=
        \frac{\text{True Positives}(\alpha)}{\text{True Positives}(\alpha) + \text{False Positives}(\alpha)}
    \end{align}
    Within everything that has been predicted as a member of cluster $\alpha$, precision counts the percentage that is correct.
    A high precision($\alpha$) indicates that although we failed to identify all members of the cluster, the ones are identified as members of $\alpha$ are very likely to be correct.
    Precision is also sometimes referred to as purity.
    \item the \emph{F1 score} is an harmonic mean of precision and recall.
    \begin{align}
        F1(\alpha) = \frac{2 \times \text{precision}(\alpha)\, \text{recall}(\alpha)}{\text{precision}(\alpha)+
        \text{recall}(\alpha)}
    \end{align}
\end{itemize}
The overall recall, precision and $F1$ score for the multi-label classification problem are computed as the weighted average over all labels $\alpha$, where the weight of each label is proportional to its occurrence in the data. 
Weighted recall is equal to accuracy. All these metrics will also be presented in percentage values.

\subsection{Model Evaluation}
When using a supervised machine learning model, it is important to assess its generalization capabilities, i.e. its ability to perform well on unseen data. The most common risk is overfitting, where the model simply memorizes the input-output mapping from the provided data which also includes noise and irrelevant features. 
This leads to poor model performance on new data, as an overfitted model would represent the underlying relationships between features and the target variable. 

Given the very small data set in hands, we use a repeated k-fold cross-validation to get a more robust assessment of the classification performance \citep{hastie2009elements}. In k-fold cross-validation, the data set is divided in $k$ subsets (folds) and the model is trained on $k-1$ of these subsets and evaluated on the remaining one. This process is repeated until every subset is used for the evaluation.
The performance of the model is then an average across these iterations. A repeated $k$-fold cross-validation, as the name suggests, is just a $k$-fold cross-validation repeated more than once. In this work, we opt for $k=3$ to balance between assessing the model's performance effectively and mitigating the variance of performance across folds.
\subsection{Classifiers}
Five different classifiers with different learning philosophies have been tested. We provide here a brief descriptions of the algorithms:
\begin{itemize}
    \item \emph{Nearest Neighbors classifier.} A $k$-neighbor classifier (\citealt{KNN}) implements learning based on the nearest $k$ neighbors of each sample: every sample is assigned to the class that has the most representatives within the nearest neighbors of the point. Rather than learning a function between inputs and outputs, this approach is more closely related to clustering algorithms.
    \item \emph{Random Forest classifier.} The Random Forest classifier (\citealt{RandomForest}; \citealt{RandomForest2}) is a type of ensemble algorithm, in which the relationships between the features and the target variable are learnt fitting multiple decision trees, each trained on a different random subsample of the data and columns. The output of the random forest is the class selected by most trees.
    It is robust to overfitting, can handle non-linearly separable data, missing data, and noisy data.
    \item \emph{XGBoost classifier.} The XGBoost classifier \citep{Chen2016xgboost} is also a type of ensemble algorithm, but
    unlike Random Forest, which trains each tree independently, XGBoost iteratively improves upon the previous models by optimizing the cumulative prediction error. This iterative process allows XGBoost to build subsequent trees to correct the errors of preceding ones.
    \item \emph{Support Vector Machine (SVM) classifier.} The SVM classifier (\citealt{SVM}) works by finding the boundary (also called the decision boundary) that best separates the data into different classes. This boundary is defined by a line (in two-dimensional data) or a hyperplane (in higher-dimensional data) that maximizes the margin between the closest data points of different classes. The SVM classifier is able to handle non-linearly separable data by transforming the data into a higher-dimensional space where a linear boundary can be found. In this work we both use a linear SVM and a radial basis function (RBF) SVM.
    \item \emph{Multi-Layer Perceptron classifier.} The Multi-Layer Perceptron (MLP, \citealt{hastie2009elements}) classifier is a kind of feedforward artificial neural network.
    The MLP classifier has the ability to learn complex, non-linear relationships between the input features and the output class. 
\end{itemize}
Hyperparameter tuning was conducted for each classifier to optimize performance on our dataset.\footnote{A hyperparameter is a parameter whose value is used to control the learning process. Consequently, it cannot be learned from the data; instead, it must be set by the user, who evaluates the machine's performance while varying its value.}

The performance of these models will be compared to two other classifiers: the \emph{dummy classifier} and the \emph{random classifier}. The dummy classifier is a model that makes predictions based on simple, pre-determined rules, such as always predicting the most common class in the data set. The random classifier is a model that randomly assigns class labels to data points without considering their features. These comparisons will help to assess if modelling the birth cluster with chemistry performs significantly better than chance.

\subsection{The feature space}\label{sec:MTL}
Regardless of the data or the model, an improperly constructed feature space will have a negative impact on the performance of the model used. The most common and natural choice is to just use as many features as possible: a first-level approach is to use all the features provided in the data set (in this case 19 chemical abundances).

This can be a sub-optimal choice because of some common difficulties arising when attempting classification in an unnecessarily high-dimensional space. These problems are well known as the ``curse of dimensionality'' \citep{BELLMAN1958}, as also discussed in the chemical tagging framework in \citep{Spina} and \citep{jackson2021using}. 

The curse of dimensionality happens because the number of samples required to accurately model the relationship between features and target variables increases exponentially with the number of dimensions. In these higher-dimensional spaces distance measures become less meaningful and thus identifying patterns more complicated. Furthermore, increasing the number of dimensions often introduces features that do not add much value to our model. When this is the case, the model may start to interpret noise or irrelevant data as significant, leading it to learn from them. This, in turn, results in a reduction in performance of the classifier.

These challenges are very relevant when it comes to using stellar chemical abundances, as they are estimated with limited precision and accuracy, and it is now believed that no more than seven to ten elements actually matter \citep{Ting_2022}. Chemical abundances are often tightly and linearly correlated, other than being noisy \citep[\textit{e.\,g.}][]{ness2018galactic, ness2022homogeneity}. Therefore, the space composed by the 19 chemical elements from our catalog very likely contains redundant features. 

Another possible limiting factor is that the chemical abundances provided in the catalog are defined using the total hydrogen content (Eq. \ref{eq:X_H}). This representation, even if commonly adopted, is just a convention. There is no evidence showing that it is the best representation for identifying co-natal stars. Indeed, using wide binaries --- which are also expected to be co-natal and chemically homogeneous such as open clusters \citep{hawkins2020chemical, espinozarojas2021consistency} showed that stars were more similar to each other in the so-called chemical clock abundance ratios discussed in \cite{jofre2021sequences} than in abundance ratios with respect to Fe. 

Overall, both the choice of the elemental abundance space and the choice of its representation are fundamental decisions that have to be taken.
That is why in this work, more than focusing on applying and evaluating what is the best classification model, we focus on the abundance space, by taking a dimensionality reduction approach.

Dimensionality reduction is the process of reducing the dimensionality of the feature space by obtaining a subset of the original features, for instance by judiciously picking a subset of the original features based on their usefulness for prediction, or by combining them to extract new ones.
This can be achieved by leveraging statistical or machine learning approaches to project raw features into desired, non-redundant features, with the goal of enhancing the relationship between features and target variables. 

This better ``perspective'' on the data may thus result in improved accuracy.
To achieve this, several approaches can be taken, among which we consider: removing the most uncertain features; removing low mutual-information score features; applying a principal component analysis; clustering correlated features together; and combining representation learning with classification via multi-task learning. These approaches are explained with more detail below.

\subsubsection{Removing noisy features}
Given the small size of our data set, during the training the model might be giving too much importance to very noisy features, thus delivering lower generalization capabilities. We can then run our tests in a data set where the most uncertain feature (Cobalt, Fig. \ref{fig:ab_uncertainties}, top panel) is removed.

\subsubsection{Removing uninformative features}
Another simple way to define a space with a smaller number of dimensions is to identify which elements are the most meaningful to predict the class. To do so, we estimate for each element the Mutual Information score \citep{MutualInfo} between every chemical abundance and the target label as
\begin{align}
    I(y;x_j)=H(y)-H(y|x_j),
\end{align}
where $I(y;x_j)$ is the mutual information for $x_j$ and $y$, $H(y)$ is the entropy for $y$ and $H(y|x_j)$ is the conditional entropy for $y$ given $x_j$.
This quantity measures the average amount of information about a random variable ($y$) that another feature ($x_j$) carries.
This score is also influenced by the intra-class entropy of each element, which in turn is influenced by the uncertainty the element is estimated with. High uncertainty means higher entropy, and higher entropy means lower Mutual Information score, regardless of the importance that feature might have to link stars to their open cluster.

\begin{figure}[tb]
    \centering
    \includegraphics[width=\linewidth]{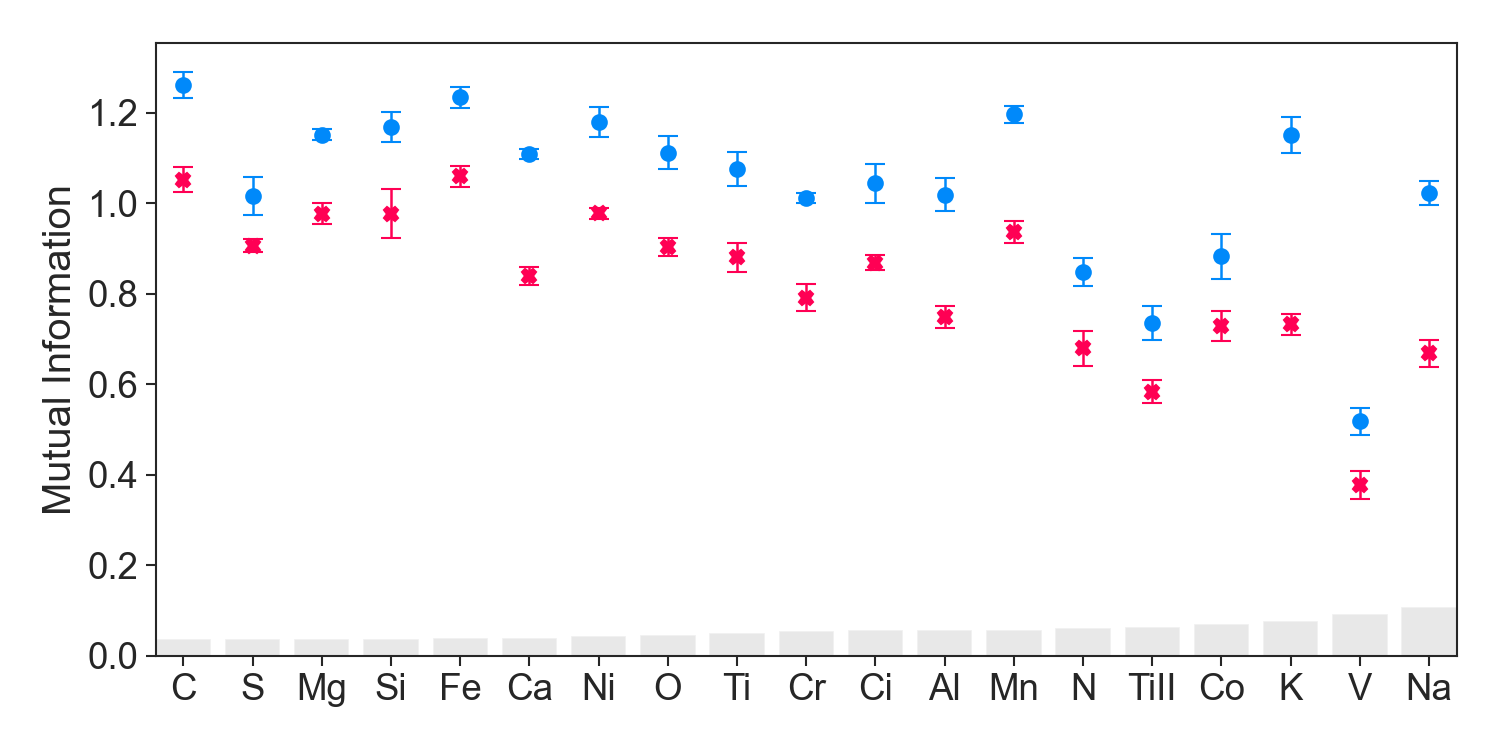}
    \caption{Mutual information (Eq. \ref{fig:MutualInfo}) between every chemical abundance ($x$-axis) and the target variable $y$. The results shown are averaged across different random subsets of the data set and the error bars indicate the standard deviation. In blue we indicate the mean chemical abundances and in red the abundances perturbed with their provided uncertainty. The grey vertical bands indicate the median error of every feature.}
    \label{fig:MutualInfo}
\end{figure}

In Fig.~\ref{fig:MutualInfo} we show in blue the mutual information scores between every chemical abundance and open cluster label and in red the ones obtained by perturbing the abundances with their provided uncertainty. 
We observe that chemical elements with lower uncertainty exhibit higher scores, with carbon being the most informative feature, and vanadium the least.

\subsubsection{Feature selection through correlation clustering}
As mentioned in Sect. \ref{sec:MTL},
many chemical abundances show strong correlations. 
The main problem caused by the presence of highly correlated features is the addition of sources of noise, without adding any new information.
To tackle this problem we performed a hierarchical clustering on the Spearman rank-order correlations, and keeping a single feature from each cluster. 

\begin{figure}[t]
    \centering
\includegraphics[width=\linewidth]{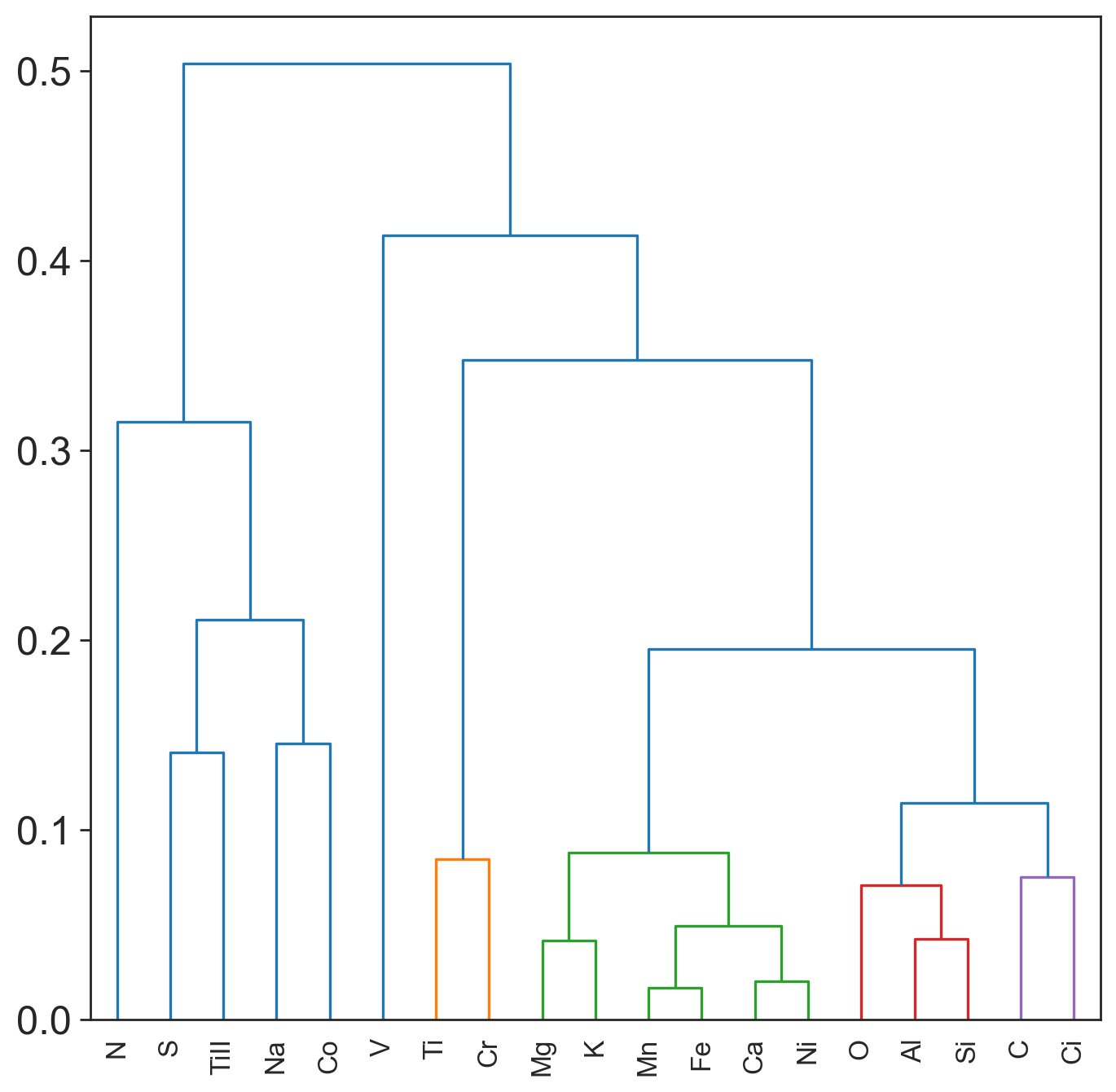}
    \caption{Hierarchical clustering on the Spearman rank-order correlations. The value of the $y$-axis is $1-$ the Spearman correlation coefficient. For example, the correlation coefficient between Ti and Cr is $\approx 0.93$. Features with a correlation  $>0.9$ are grouped together, as indicated by the branch colors.}
\label{fig:features_correlation}
\end{figure}

In Fig.~\ref{fig:features_correlation} we show a dendrogram representing the arrangement of the elements that have been clustered based on their Spearman order correlation.
Here each tip represents a chemical element, and the length of the branches represents the correlation distance between features. Elements that are more similar to each other are joined together under shorter branches, while those that are less similar are joined under longer branches.

As expected, the correlations follow the nucleosynthetic origins of the elements \citep{Kobayashi2020origin}. For example, Mn and Fe, both iron-peak elements, exhibit the strongest correlation. Similarly, elements such as O, Al, Si, and C/Ci, $\alpha$-elements, display significant collinearity.
Interestingly, the feature least correlated with all the others, V, is also the feature that carries the least information about $y$ (Fig. \ref{fig:MutualInfo}).
\subsubsection{Principal component analysis}
The most popular dimensionality reduction technique is the Principal Component Analysis (PCA, \citealt{PCA}), which has been extensively discussed in the context of chemical tagging (\textit{e.g. }\citealt{ting2012principal, BlancoCuaresma2015testing}). 
PCA consists in linearly rotating the data (with $d$ features) into a new coordinate system where most of the variation in the data can be described with $d'<d$ dimensions (principal components).
Additionally, a beneficial byproduct of this process is the expected reduction in noise.

\begin{figure}
    \centering
    \includegraphics[width=\linewidth]{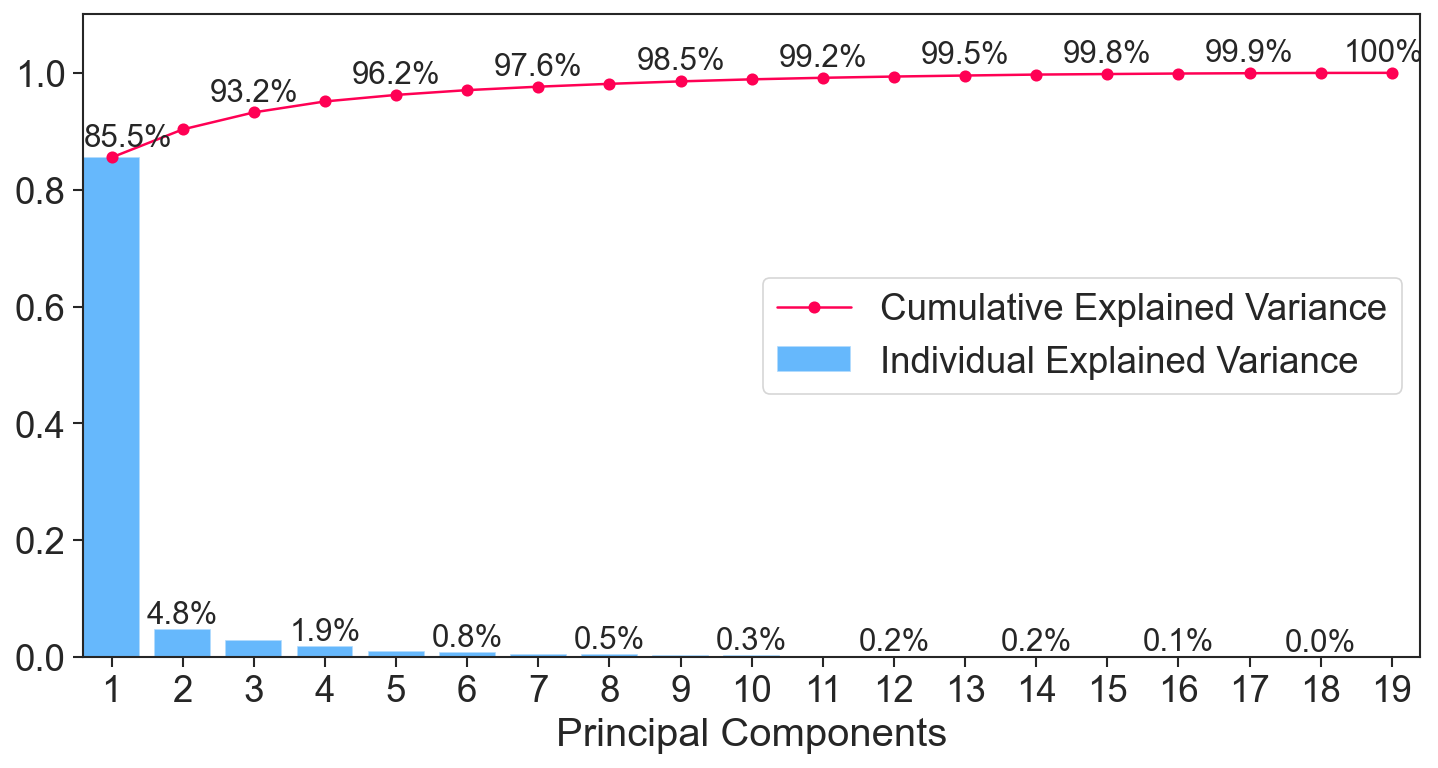}
    \caption{Individual and cumulative explained variance by principal components. The blue bars represent the percentage of variance explained by each individual principal component, while the red line indicates the cumulative explained variance. This plot demonstrates that $85.5\%$ the total variance in the dataset is captured by only one component, and that only ten components are required in order to capture more than $99\%$ of the total variance.}
    \label{fig:pca_exp_var}
\end{figure}

Although finding the direction of greatest variation in a data set may seem like a good approach, it does not always result in greater separation of groups in the transformed space and improved results. In fact, it may have the opposite effect and blur the distinctions between classes \citep[\textit{e.\,g.}][]{zheng2021application}. In Fig.~\ref{fig:pca_exp_var} we show the variance explained by each principal component with blue bars, and the cumulative explained variance in red. 
This demonstrates that more than $85\%$ of the variation in our dataset can be captured by only one component and that only ten of them contain more than $99\%$ of the total variance.
\subsubsection{Multitask Learning}

Multitask learning \citep{thrun-1995:multi,caruana-1997:multitask,Crawshaw2020} is a machine learning technique that uses a single model to perform multiple related tasks. It aims to improve the performance of related tasks by consolidating information from them in a single representation. This leads to several advantages, including improved data efficiency, faster model convergence and reduced model overfitting. It is also known as joint learning, learning to learn, and learning with auxiliary tasks.

In our case, the goal is to learn both the representation and the classification tasks simultaneously. Rather than training independent models for each of these tasks, we allow a single model to learn to complete them at once. In this process, the model uses all of the available data information to learn representations of the data that are useful for linking chemical abundances to birth clusters.

The motivation behind using multitask learning is to identify the optimal lower dimensional representation of the input features, by leveraging the knowledge of abundance uncertainties and incorporating field stars data from APOGEE to enhance and generalize the representation learning process.

\begin{figure}[tb]
    \centering
    \includegraphics[width=0.9\linewidth]{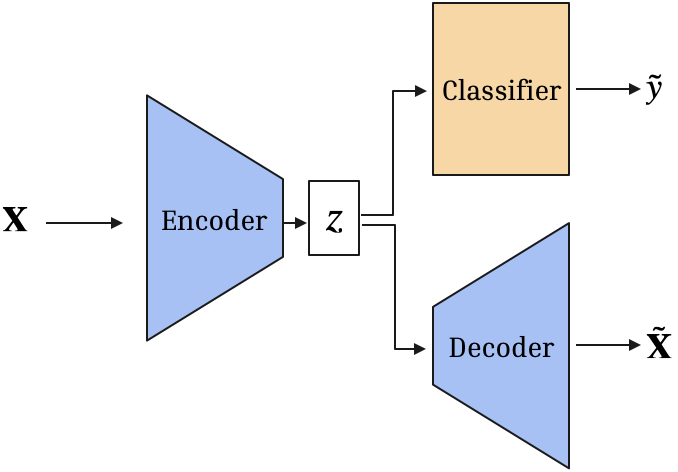}
    \caption{Representation learning architecture used in this work. The feature extraction task is assigned to the autoencoder (constituted by encoder and decoder) to learn more clean data and more efficient features, while the classifier classifies and regularizes this new representation.}
    \label{fig:ae_clf}
\end{figure}

Several multi-task models were thus tried and tested on the same data set. The most representative one is shown in Fig.~\ref{fig:ae_clf}.
It is composed by a one layer encoder, followed by a decoder of the same size. To the latent space we append a single fully connected layer with softmax activation function, acting as a classifier. In this way, we regularize the latent space by forcing the encoder to represent classes in a linearly-separable way. This simple classification module will also ease the training process, likely reduce overfitting and possibly improve interpretability.

One can think of it as a Multi-Layer Perceptron classifier, but the reconstruction auxiliary task explicitly forces the network to learn an encoder that produces a representation that is informative enough to be able to reconstruct the original data.

To optimize this model, we define the loss function
\begin{equation}
    L=L_{\rm rec}+\gamma L_{\rm class}\,,
\end{equation}
where $L_{\rm class}$ is the classification loss defined as the sparse categorical cross entropy
\begin{equation}
    L_{\text{class}} = -\sum_{i=1}^m\sum_{l=1}^n y_{il}\log(p_{il})\,,
\end{equation}
where $y_{il}$ is the true label of the sample $i$ and $p_{il}$ is the predicted probability for it belonging to the $l^{th}$ class.

As our data contain noise, mostly caused by observational Gaussian noise in the spectra and uncertainties in the abundances estimations pipelines, we train the autoencoder by optimizing a weighted version of the mean squared error
\begin{equation}
    L_{\rm rec} = \frac{1}{m}\sum_{i=1}^m\sum_{j=1}^d(x_{ij}-\Tilde{x}_{ij})^2e^{-\sigma_{ij}}\,,
\end{equation}
where $x_{ij}$ and $\Tilde{x}_{ij}$ are the data and the reconstructed $j$-th abundance of star $i$, respectively, and $\sigma_{ij}$ its reported uncertainty. 
The goal is to learn clean data from noisy data to promote the robustness of the algorithm in the classification task.
\section{Results}\label{sec:results}
In Table \ref{table:baseline_perf} we report the classification performance, obtained using different classifiers and the feature space defined by all the 19 chemical features provided in the catalog.
\begin{table}[tb]
\centering       
\caption{Classification performance.}
\begin{tabular}{lcccc}          
\toprule          
\textbf{Classifier} & \textbf{Accuracy} & \textbf{Precision} & \textbf{F1 score}\\
\midrule
Dummy Classifier & $20.8\pm0.2$ & $83.5\pm0.1$ & $7.1\pm0.1$ \\
Random Classifier & $3.5\pm0.6$ & $7.5\pm1.9$ & $31.7\pm6.8$ \\
Nearest Neighbors & $66.0\pm2.0$ & $66.3\pm2.5$ & $70.8\pm2.6$ \\
XGBoost & $66.4\pm1.9$ & $66.5\pm2.2$ & $70.5\pm2.6$ \\
MLP & $67.1\pm2.3$ & $68.8\pm1.9$ & $71.3\pm2.2$ \\
Random Forest & $67.8\pm2.1$ & $67.7\pm2.5$ & $71.1\pm2.4$ \\
Linear SVM & $69.7\pm1.9$ & $69.8\pm2.0$ & $72.7\pm2.3$ \\
RBF SVM & $70.1\pm2.0$ & $70.6\pm2.2$ & $72.8\pm2.1$ \\
\bottomrule
\end{tabular}
\tablefoot{Classification performance (expressed as percentages), averaged across folds, obtained with different classifiers using all the 19 abundances provided in the catalog.}
\label{table:baseline_perf}
\end{table}
The highest metrics achieved among the classifiers, although being significantly better than random chance, are still around $70\%$. This performance may be considered low in the context of the classification task at hand. 
What we also note here is how the performance obtained does not change significantly with the classifier used. It is then possible to conclude that the bottleneck in the performance is not how we chose to model the data, but the data itself.
We also note that the top-performing models in this study, as indicated by highest metrics and lowest dispersion, are the two support vector machines, the less complex and less parametric models (after the nearest neighbour classifier) tested.

Given the similar behavior shown by all the classifiers tested, we retrain a RBF SVM classifier, always with a repeated 3-fold cross validation, and for every iteration compute the recall and precision for every cluster. The results reported afterwards refer to the average of these quantities. We use them as an example to understand our results further.

\subsection{Why are some clusters not recovered? - Looking into the confusion}\label{sec:confusion}
\begin{figure*}[tb]
    \centering
    \includegraphics[width=\linewidth]{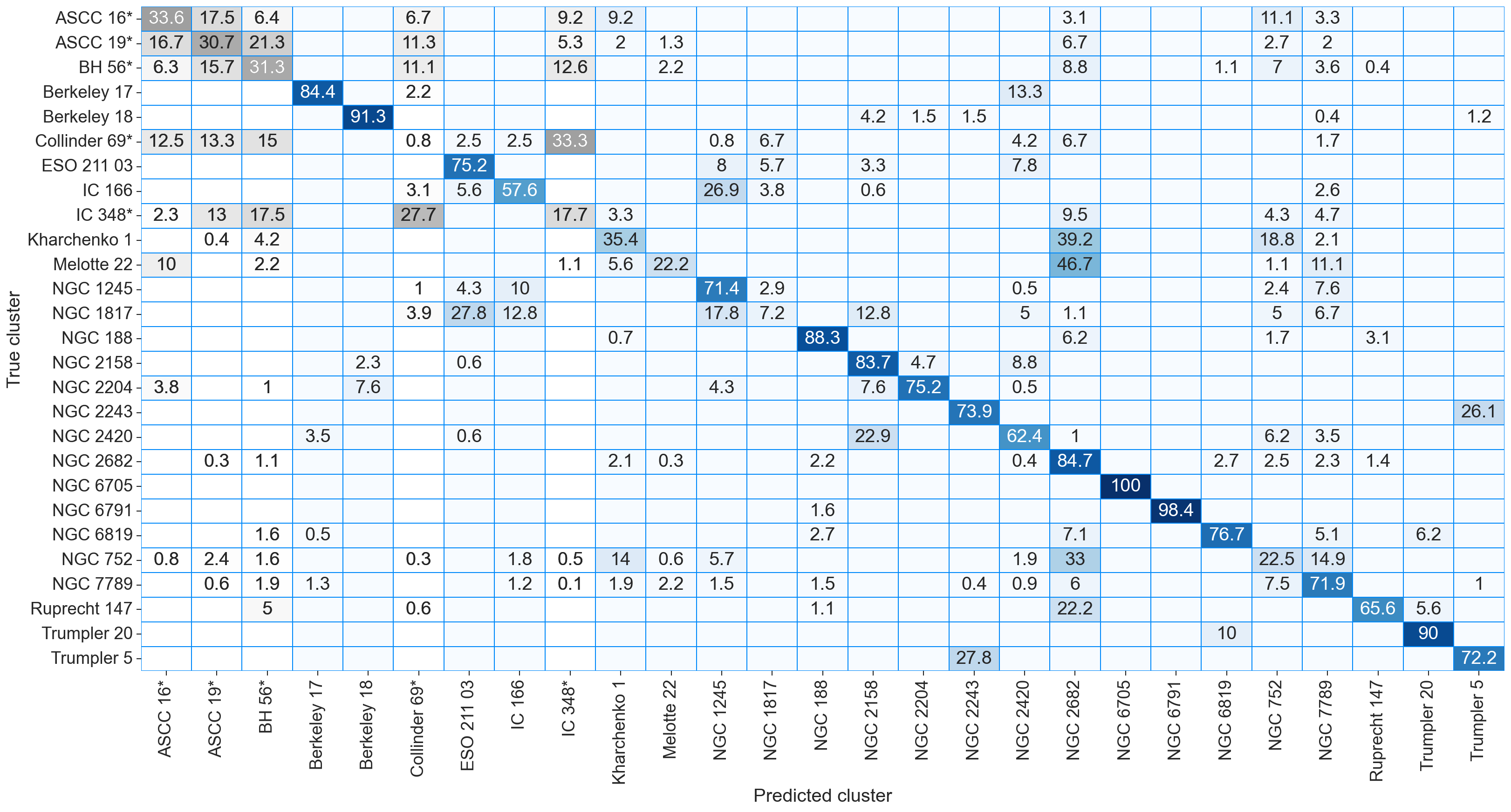}
    \caption{Average over folds confusion matrix for clusters members classification. Each row of the matrix represents the percentage of samples in an actual cluster, while each column represents the percentage of samples predicted in a cluster. Thus, the normalization of the matrix is such that the sum over columns is 100. Cells with missing annotation contain zero samples. 
    For example, considering the second to last row (Trumpler 20): on average, $90\%$ of members are correctly classified as such, while $10\%$ as members of NGC 6819.
    Rows and columns in shades of gray indicate clusters not flagged as high-quality.}
    \label{fig:confusion}
\end{figure*}

In Fig.~\ref{fig:confusion} we report the average (over folds) confusion matrix. The diagonal entries represent the percentage of correctly classified samples (i.e. the classifier's recall per cluster), while off-diagonal entries indicate misclassifications.

Generally, we can observe some clusters, like NGC 6705 and NGC 6791, are always recovered (almost 100\% recall, very high precision), while others, for e.g. Collinder 69, are never recovered (very low recall). We further observe that most of the largest confusions are between ASCC 16, ASCC 19, BH 56, Collinder 69 and IC 348 (see, for example, Fig.~\ref{fig:confusion}, first row). These clusters are all flagged as not high-quality clusters in the OCCAM catalog. 
Finally, we find that some members belonging, for example, to Kharchenko 1 or NGC 752, are assigned to  NGC 2682, the class with most members in our data set (see Table~\ref{table:counts}).
The non-recovery of certain clusters can be easily attributed to their inherently distinct and non-unique chemical patterns. This would clearly make impossible to definitively link specific chemical signatures to particular birth clusters, regardless of the methods and quality of data in hand. 
Our objective is to explore alternative factors contributing to this confusion and determine if resolving these issues can facilitate the establishment of this link.

We identify four main factors within the data set in use: the quality of the clusters, the imbalance between classes, the uncertainties associated with abundances, and the biased importance attributed to certain features. These factors are discussed further below.
\begin{figure*}[tb]
    \centering
    \includegraphics[width=\linewidth]{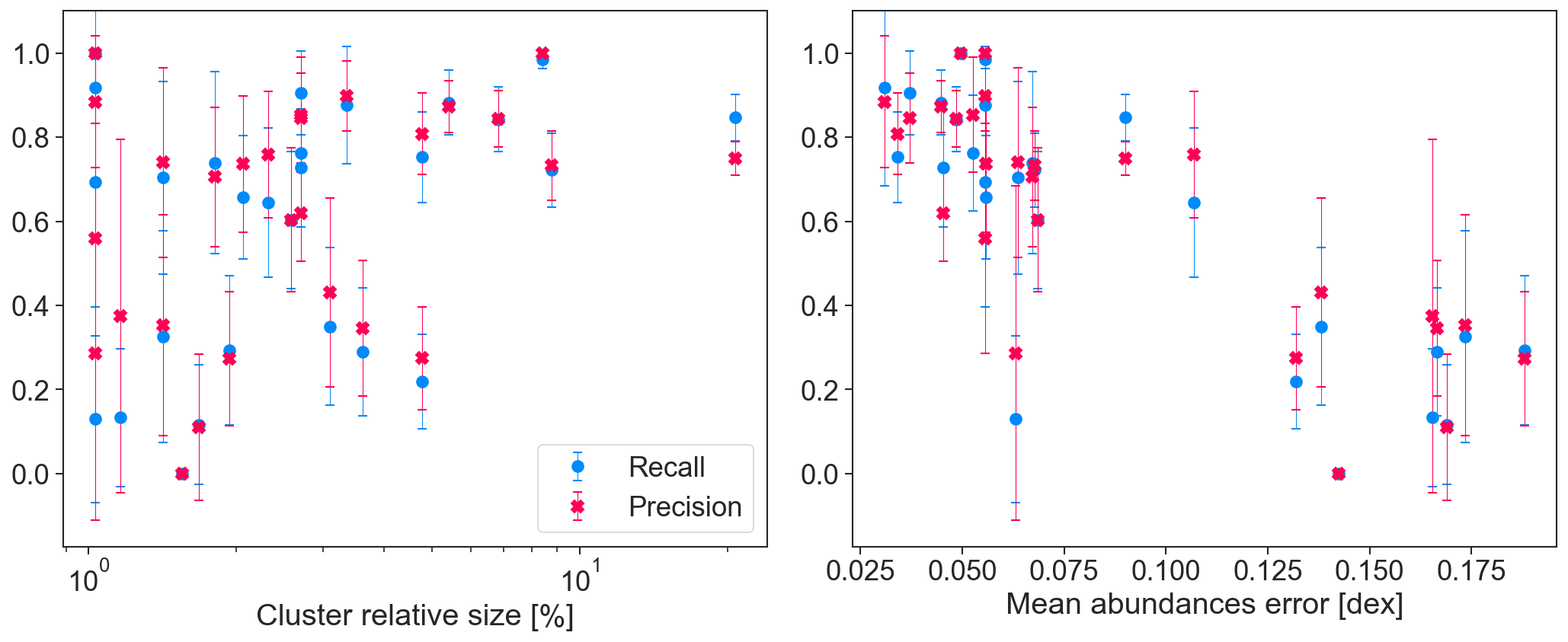}
    \caption{Average over folds recall (blue) and precision (red) as function of (left) size of cluster and (right) average over cluster members of chemical abundances uncertainties. 
    Each point represents a high-quality flagged open cluster, with error bars indicating the standard deviation across folds/repetitions. Clearly, the uncertainty decreases when the cluster size increases due to sampling effects.}
    \label{fig:perf_trends}
\end{figure*}
\begin{table}[tb]
\centering              
\caption{Spearman correlation coefficient $\rho$ between data properties and performance.}
\begin{tabular}{r c c | c c}
\toprule
\multirow{2}{*}{}&
\multicolumn{2}{c}{\textbf{Recall}}&\multicolumn{2}{c}{\textbf{Precision}}\\
&$\rho$  & $p$-value & $\rho$  & $p$-value \\
\midrule
Cluster size & 0.30&   0.12& 
        0.27&0.17 \\
$\langle\sigma\rangle$ & $-$0.79 &   $<2\times10^{-6}$& $-$0.77& $<3\times10^{-6}$ \\
$\left[\text{Fe/H}\right]$ & 0.38 &   0.05 & 0.37 & 0.05\\
\bottomrule
\end{tabular}
\tablefoot{Spearman correlation coefficient $\rho$ and its corresponding p-value between cluster relative size, mean abundance error and mean metallicity of every cluster and recall and precision. The p-value roughly indicates the probability of an uncorrelated system producing datasets that have a Spearman correlation at least as extreme as the one computed from these datasets.}
\label{table:spearman}
\end{table}

\paragraph{Quality of clusters:} 
We first evaluate how the cluster quality flag provided in the OCCAM catalog affects the quality of classification. This flag is based on a visual assessment that evaluates whether stars that pass the combined radial velocity, proper motion and metallicity criteria also lie in a sensible position in the observed cluster color-magnitude diagram (CMD), considering their spectroscopically determined $\log{g}$.  Clusters with poorly populated main sequences, or significant contamination from field stars are flagged as ``potentially unreliable''. 
For more details refer to \cite{OCCAM2}.

From Fig.~\ref{fig:confusion} we see how none of the clusters flagged as ``potentially unreliable'' is decently recovered, as can also be seen from Fig.~\ref{fig:clusters_perf_distribution}. Considering only high-quality clusters, the performance for all the classifiers improve. In particular for an RBF SVM the accuracy, precision and F1 score are, respectively, $75.2\pm2.3$, $75.5\pm2.3$, and $75.7\pm2.3$.

We thus decide to remove these clusters from the following analyses, as their assigned members are not very likely to be actual birth siblings. We further note that the stars belonging to the clusters flagged as potentially unreliable are predominantly cool dwarfs and pre-main sequence stars. This poses further uncertainties in their parameters and abundances.  As a result, clusters with poor quality have larger uncertainties and abundance spread (see Fig. \ref{fig:ab_uncertainties}). The spectral analysis of cool dwarfs and pre-main sequence stars was not optimized for ASPCAP thus atmospheric parameters and abundances are more uncertain for such stars \citep[see extensive discussions in][]{Jonsson20}. Since the training set used for AstroNN, which is the catalogue we use in this work, is based on ASPCAP catalogues, we expect these labels to be more uncertain.  We remark however that the scope of this paper is to assess how with state-of-the-art data analysis methods can we retrieve open clusters as provided in public catalogues.


\paragraph{Imbalanced data set:}
Imbalanced classification \citep{Krawczyk2016learning} refers to a scenario where the distribution of classes in the target variable is unequal, i.e. when the number of samples in one or more classes (minority classes) is significantly lower than other classes (majority classes). This imbalance can result in models that have poor predictive performance for the open clusters with less members in the training data set.
We display the trend of average recall in blue and precision in red for each class in relation to their size in the left panel of Fig. \ref{fig:perf_trends}. We can see how there is no notable difference in performance between small and large classes hence rejecting the size of the class as a source of uncertainty in the results.

\paragraph{Abundances uncertainties:}
Generally speaking, when there is strong noise, the performance of machine learning methods decrease.
In the right panel of Fig. \ref{fig:perf_trends} we show in blue the recall and in red the precision as function of the average abundance error. We can see a clear negative trend of both these metrics. The Spearman correlation coefficient for the average abundance uncertainty for both the recall and precision is $\approx -0.78$ with a low $p$-value. This shows that the uncertainty in the measurements have a significant impact in the recovery of clusters.

\paragraph{Biased Feature Importance:}
Especially given the small size of our data set, during the training the model might be giving too much importance to noisy and/or irrelevant features, thus delivering lower generalization capabilities. We test this source of discrepancy by applying the feature selection and extraction methods described in Sect. \ref{sec:MTL}. The results are reported and discussed in the next sections.

\subsection{Changing the chemical representation}
Motivated by the performance trends as function of the chemical abundances uncertainties see right panel of Fig. \ref{fig:perf_trends}), we run the same analysis as before, but with the different representations of the input features described in Sect. \ref{sec:MTL}. By reducing the number of input features, we hope to decrease the noise in the data and thus improve the performance, or at least obtain the same results as given by using all the chemical abundances provided.

For this analysis, we maintained the same hyperparameters that were previously optimized for the full feature set, to establish a consistent baseline for the comparison of performance across the various representations. 
By doing so, we aim to isolate the impact of chemical representation on model efficacy.

In Table \ref{table:total_res} we report the classification performance from a RBF SVM classifier, using only high-quality clusters and different input feature spaces. The values are sorted in increasing accuracy.
\begin{table*}[tb]
\centering                  
\caption{Classification performance obtained with different input feature spaces.}
\begin{tabular}{lcccc}          
\toprule          
\textbf{Input Features} & \textbf{Accuracy} & \textbf{Precision} & \textbf{F1 score}\\
\midrule
Clustered abundances & $72.4\pm2.1$ & $72.7\pm2.2$ & $73.2\pm2.4$ \\
10D PCA & $73.3\pm2.3$ & $73.6\pm2.4$ & $73.6\pm2.4$ \\
10D MTL & $73.8\pm2.4$ & $74.0\pm2.5$ & $73.8\pm2.4$ \\
9D PCA without V & $74.1\pm2.3$ & $74.2\pm2.5$ & $74.1\pm2.4$ \\
Abundances without Co & $75.0\pm2.1$ & $75.3\pm1.9$ & $75.4\pm2.3$ \\
\hline
All abundances & $75.2\pm2.3$& $75.5\pm2.3$& $75.7\pm2.3$\\
\hline
Abundances without V, Co, TiII & $76.0\pm2.3$ & $76.5\pm2.4$ & $75.5\pm2.2$ \\
Abundances without V, Co & $76.1\pm2.2$ & $76.8\pm2.3$ & $75.7\pm2.2$ \\
Abundances without V, TiII & $76.2\pm2.2$ & $76.6\pm2.3$ & $76.0\pm2.4$ \\
Abundances without V & $76.4\pm2.4$ & $76.7\pm2.5$ & $76.5\pm2.4$ \\
\bottomrule

\end{tabular}
\tablefoot{Classification performance (expressed as percentages), averaged across folds, obtained with different input feature spaces, using only high-quality clusters and a RBF SVM classifier. Clustered abundances refers to the subset of chemical elements defined through correlation clustering: Si, C, Mg, Fe, Ti, S, Na, N, V, TiII (see Fig. \ref{fig:features_correlation}).}
\label{table:total_res}
\end{table*}
From the Table we find that removing the feature with the lowest mutual information score (vanadium), led to a slight improvement in the performance of the classifier. This finding underscores the risk of overfitting in our simple models. The feature space defined through this process yields the best average performance in this work. We further find that removing the most noisy feature (cobalt) and the two with the lowest information score (vanadium and titanium II), we also obtain an improvement in the performance.

It is worth to comment that the use of PCA as a feature reduction technique, even if applied on a data set without nuisance features like V, resulted in a poorer performance compared to the baseline.  As anticipated in Sect. \ref{sec:MTL}, while PCA efficiently condenses feature space by capturing maximum variance, it may not necessarily preserve the discriminative information for our task of distinguishing between open clusters.
\subsection{On the memberships selection criteria}
The reliability of the assigned membership probabilities is another crucial factor that could limit performance. To assess the robustness of our findings to variations in membership probability selection, we replicated the main analyses considering as members of open clusters all the stars with $p_{\rm RV}>0.6$, $p_{\rm Hunt}>0.9$. $p_{\rm Hunt}$
indicates probabilities from \cite{Hunt2023improving}, computed for sources in Gaia DR3 \citep{Vallenari2023gaiadr3}
down to magnitude $G\sim20$ using 
the HDBSCAN clustering algorithm.

Results remained qualitatively consistent, exhibiting a comparable range of accuracy and precision scores (the best model being an SVM classifier, with accuracy and precision  of $70.8\%$ and $72.9\%$, respectively).
The most influential factors for successful cluster recovery continued to be the precision of chemical abundances and the quality of clusters (see Sect. \ref{sec:confusion}). Also, the behavior with different feature spaces remained consistent. 
We therefore argue that despite potential slight quantitative differences arising from different membership probability choices, the core results presented in the previous sections remains unaltered.
\subsection{Adding random field stars}
It is important to include field stars in the evaluation of chemical tagging because they provide a representative sample of the overall population of stars in the Milky Way and is the ultimate dataset for which chemical tagging aims to be applied \citep{Freeman2002newgalaxy}. This allow us to validate if we can still identify birth siblings in a training set highly contaminated by stars in the general field, providing a critical benchmark for evaluating the feasibility of strong chemical tagging.
To do so, we repeat the analysis presented in Sect. \ref{sec:methods}, with only high-quality clusters with at least 12 members, all the 19 chemical abundances and an RBF SVM classifier with a training set with varying number of field stars categorized as ‘‘Field’’, whose chemical abundances are always retrieved from the AstroNN catalog for APOGEE DR17.
These field stars are sampled randomly from this catalog and are required to have a metallicity between $-0.6$ and $0.3$, a range similar to that of our open clusters (see Table \ref{table:counts}).

We train the classifier in a balanced way, i.e. the weight assigned to each sample is adjusted to be inversely proportional to the frequency of its respective class\footnote{The weight assigned to members of class $\alpha$ is $N_{\rm samples} / (N_{\rm classes} \times N_{\alpha})$, where $N_{\rm samples}$ is the total length of the training set, $N_{\rm classes}$ is the number of classes and $N_{\alpha}$ the number of sample belonging to class $\alpha$.}. This assigns equal importance to all classes present in the data set.

\begin{figure}[h!]
    \centering
    \includegraphics[width=\linewidth]{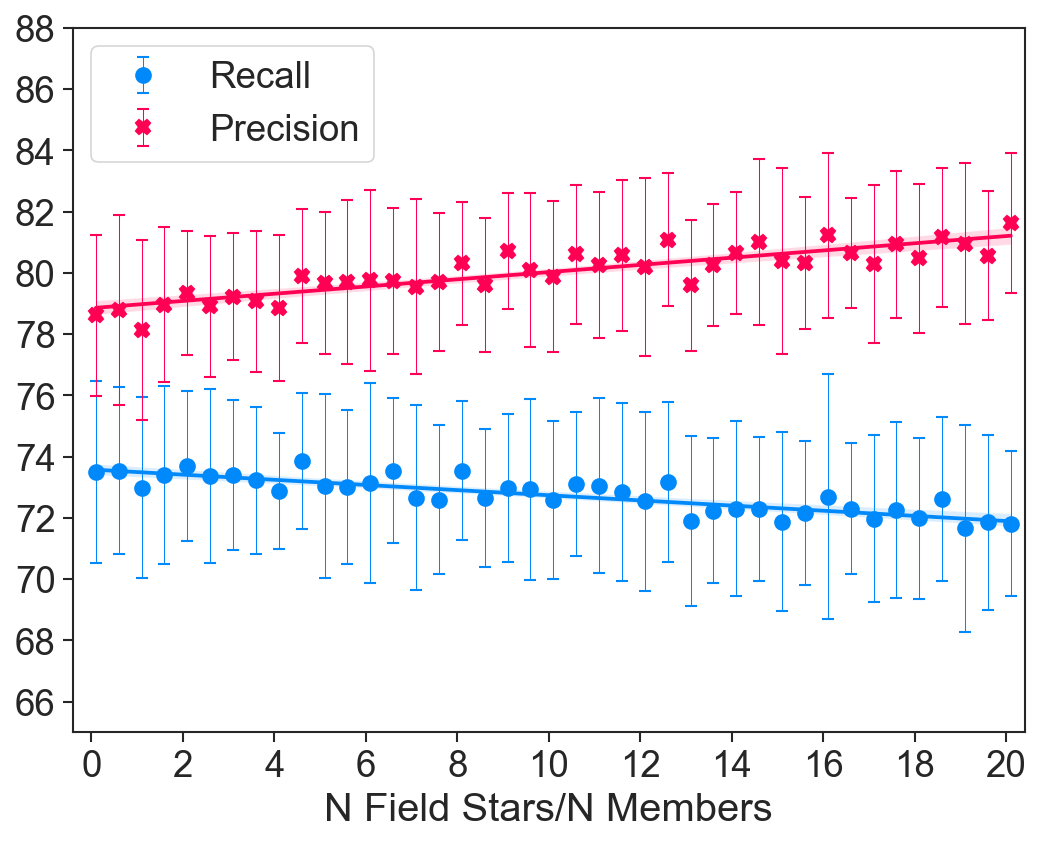}
    \caption{Recall (blue) and precision (red) as function of the ratio between the number of field stars and number of cluster members in the train set. The corresponding linear fits are also shown.}
    \label{fig:contaminants}
\end{figure}

We report in Fig. \ref{fig:contaminants} the recall and precision computed only for cluster members (i.e. not field stars) as function of the ratio between the number of field stars and number of cluster members in the train set.
We observe a decrease in recall as the ratio of field stars to cluster members increased in the training set. This shows how the contamination introduced by field stars, makes the task much more complex, even in the supervised scenario.
On the other hand, the precision increases as the number of field stars increases: if the model predicts that a star belongs to an open cluster, the probability of this prediction being correct is high when precision is high. Essentially, a higher number of stars are classified as field stars.

To further demonstrate how the inclusion of random field stars affects the classification, we found that for one of the best-recovered open clusters, NGC 6791, the recall goes from $98.4\%$ (see Fig. \ref{fig:confusion}, seventh to last row) to $\approx 82\%$ when the field stars are ten times the members in the training set, as $18\%$ of its members are classified as field stars.

It is worth to consider the possibility that some of the field stars added in the dataset might indeed be siblings to the open clusters members which left the cluster in the past. Open clusters evaporate and populate the field with their members \citep{Krumholz2019starclusters}. We did not perform an evaluation to know which of the field stars could have been indeed members of the open clusters. 
\section{Summary and discussion}\label{sec:discussion}
In this study we have attempted a supervised approach to determine if chemical abundances alone can provide enough information to allow us to identify co-natal stars that have dispersed in the Milky Way, and thus enable the reconstruction of the star-formation history of the galaxy. 
To do so, we considered chemical abundances of a sample of open cluster stars which we know they are co-natal. 
Our work was motivated by the recent findings that have shown that such stellar groups are hard to recover via chemistry alone in all spectroscopic surveys (e.g. \citealt{Mitschang2014quantitative, ness2018galactic, Price-Jones2020, Casamiquela_2021}). Given that current methods to analyze on-going large spectroscopic surveys are still not able to estimate both accurate and precise chemical abundances for a large sample of cluster members, recovering open cluster members from clustering chemistry alone remains a hard task. This raises questions about the feasibility of strong chemical tagging, at least in this era. To answer that question, we first should evaluate if a discriminative relationship between chemical abundances and birth stellar associations is there.

The goal of this paper was thus to assess the existence of this relationship using supervised learning and, if found, to identify a better representation space for chemical abundances to recover the birth associations of known open cluster members in APOGEE DR17. 
This approach is meant to provide a baseline for evaluating the relationship between chemical patterns and birth stellar cluster and thus setting un upper limit in our hopes to perform strong chemical tagging.
We summarize our main analyses below.
\begin{itemize}
    \item We applied various several machine supervised learning techniques (specifically Nearest Neighbors, Random Forest, XGBoost, Support Vector Machines and Multi-Layer Perceptrons) for the classification of open clusters based on all 19 chemical abundances provided in AstroNN (exluding phosphorous), finding these features carry a certain amount of information about the cluster membership of stars. In particular, considering only high-quality clusters, we obtain an accuracy, a precision and an F1 score, respectively, of $75.2\pm2.3$, $75.5\pm2.3$, and $75.7\pm2.3$\\
    These percentages, although significantly better than the dummy and the random classifier, do not significantly change by using the different machine learning techniques, making us conclude that the main influence in the relationship lies in the features and not in the model employed.
    \item   This finding is further supported by the evident correlation between performance and the average abundance error we found.  We did not find a correlation between performance and cluster size, but found that for low-quality clusters the results were more uncertain. This underscores the importance of data quality and feature engineering. 
    \item We applied various feature selection and engineering techniques, finding that there is room for improving the feature space, even in a supervised scenario. For instance, the simple removal of a less informative feature vanadium led to enhanced model performance. We found that the best performance is obtained by removing V from the input chemical features, reaching a $76.4\%$ accuracy, $76.7\%$ precision under the RBF SVM classifier.     
    \item We determined precision and recall for open clusters while adding a different the number of field stars to the data set, obtaining a decline in recall as the number of field stars increases. This indicates the high difficulty in identifying birth siblings within this contaminated scenario.
\end{itemize}
With these results, we address the specific questions posed in Sect. \ref{sec:intro} as motivation of this work.

\textit{Is there actually a link between chemical abundances and birth open clusters?}
While there is a discernible link between chemical abundances and birth open clusters, this relationship is not strong enough to reliably recover disrupted clusters using chemical abundances alone. Even under a supervised approach, we found that we are not able recover open clusters with satisfying performance: we are not able to fully identify which stars belong to which cluster using chemistry alone, even using 19 chemical species obtained from high-resolution spectra. This is further emphasised with the inclusion of field stars.
While certain clusters, like NGC 6791, are consistently identified, their distinctive chemical compositions amidst the field stars categorize their identification more as an anomaly detection task \citep{Chandola2009anomaly} rather than conventional classification. 

\textit{What are the most important factors of variation in the recovery of open clusters?} Our findings clearly indicate that the precision of chemical abundances, coupled with the robustness of cluster membership determinations, are the most influential factors in the successful recovery of open clusters.

\textit{Which are the most important chemical elements? Is it possible to find a better representation for these chemical abundances in a physics-agnostic way?}
While we didn't find elements significantly more important than others, we found that the inclusion of elements like V and TiII was detrimental.
This highlights the non-trivial nature of finding the best feature space for this kind of problem, but it also stresses the potential for optimizing feature sets to improve results. The impact of feature selection and engineering is particularly pronounced in clustering analysis, where the choice and quality of features are critical due to the need for meaningful distance metrics.

Overall, given the superior capabilities of classification compared to clustering analysis --namely its higher accuracy, precision, and ability to handle complex data patterns-- and recognizing that strong chemical tagging aims to recovering disrupted clusters without explicit labels, these findings collectively challenge the feasibility of strong chemical tagging with the current measurements of chemical abundances. 

Using a supervised machine learning approach introduces certain limitations when compared to the more common clustering methods applied for chemical tagging. While unsupervised methods identify groups in the data without being influenced by predefined labels, supervised models rely on labeled data to train, making them heavily dependent on the quantity of ground truth labels.
In this study, the limited number of stars available for supervised learning analysis, particularly in certain clusters, introduces some statistical uncertainties and limits the generalizability of our findings. 
The availability of a larger sample of stars representing various birth clusters could potentially alleviate this issue.

The choice of a catalog or of the selection criteria may also lead to potential biases that can impact the reliability of the analysis. Findings derived from a specific survey, such as APOGEE DR17, might not directly translate to other surveys characterized by distinct instrumentation, data quality, or target and feature selection strategies. Consequently, the findings may carry survey-specific characteristics.
To mitigate this, future research could involve applying the same approach to other surveys, for broader validation and comparability of results (see recent discussion of \citealt{manea2023chemical} about doppelg\"angers in different surveys).

The fact that the performance of our classification is reduced when field stars are included in our dataset challenges the feasibility of chemical tagging even more. Considering that the main purpose of strong chemical tagging is to reconstruct dissolved clusters from the field, and so identify the building blocks of the galaxy, a promising research direction would involve incorporating supplementary features beyond chemistry, such as kinematical or age information. This augmentation might enhance the identification performance of birth stellar clusters, presenting an opportunity for advancing research in this field.
\begin{acknowledgements}
TS acknowledges support from Inria Chile and Millennium Nucleus ERIS NCN2021\_017, Centros ANID Iniciativa Milenio. PJ acknowledges partial financial support of FONDECYT Regular Grant Number 1231057. TS expresses his gratitude to Payel Das for initial guidance and support.
\end{acknowledgements}
\bibliographystyle{aa}
\bibliography{references}
\begin{appendix}
\onecolumn
\section{Additional figures}
\begin{figure}[ht]
   \centering
   \includegraphics[width=\linewidth]{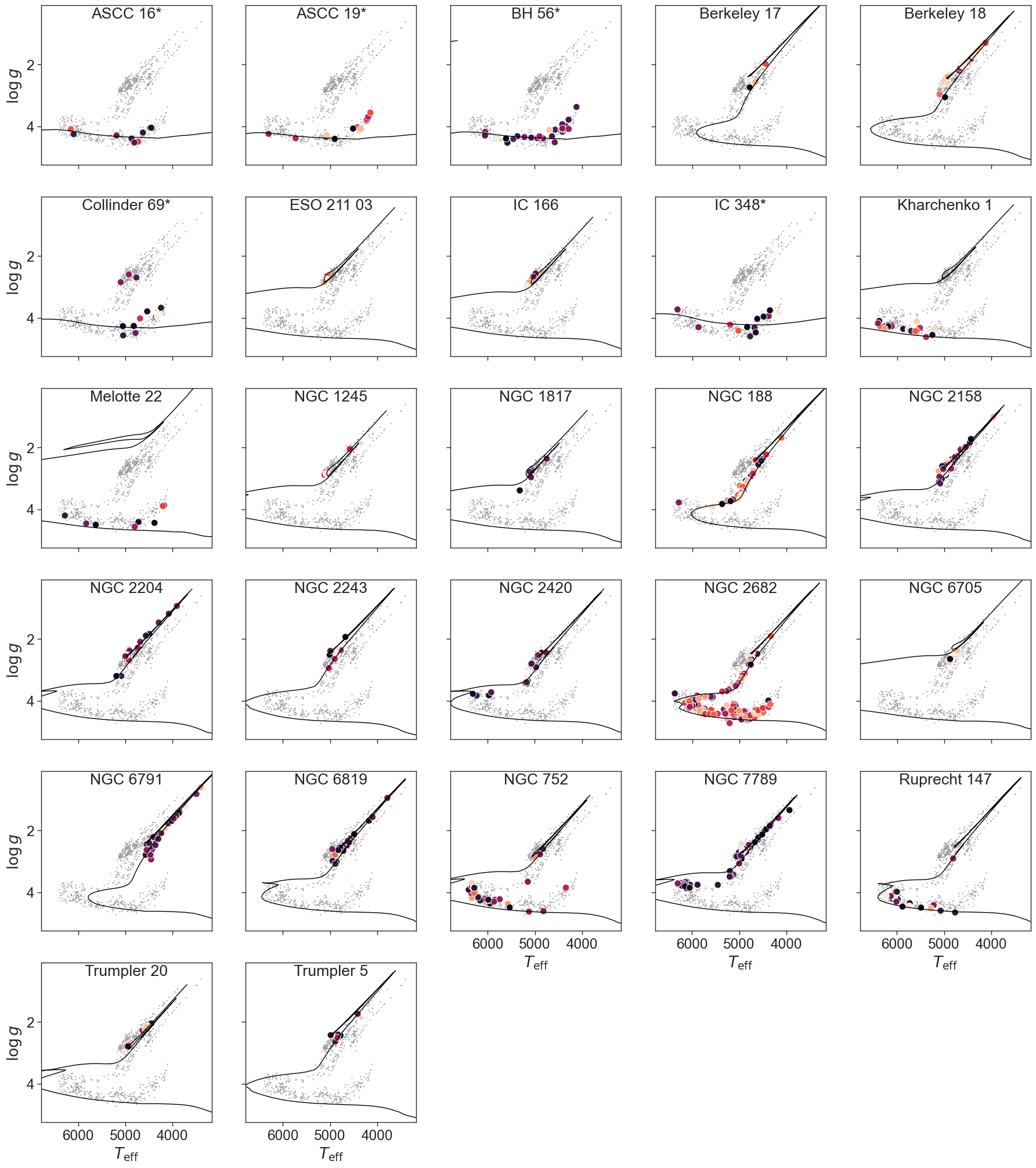} 
   \label{fig:HR}
   \caption{Kiel diagrams for open clusters members used in this work, color-coded by $p_{\rm RV}\times p_{\rm PM}$, with lighter colors indicating higher probabilities. We also overlay the relative {\tt PARSEC} isochrones \citep{bressan2012parsec}.  Grey stars in the background represent the whole sample. (*) clusters flagged not as high-quality.}
\end{figure}
\begin{figure}[ht]
    \centering
    \includegraphics[width=\linewidth]{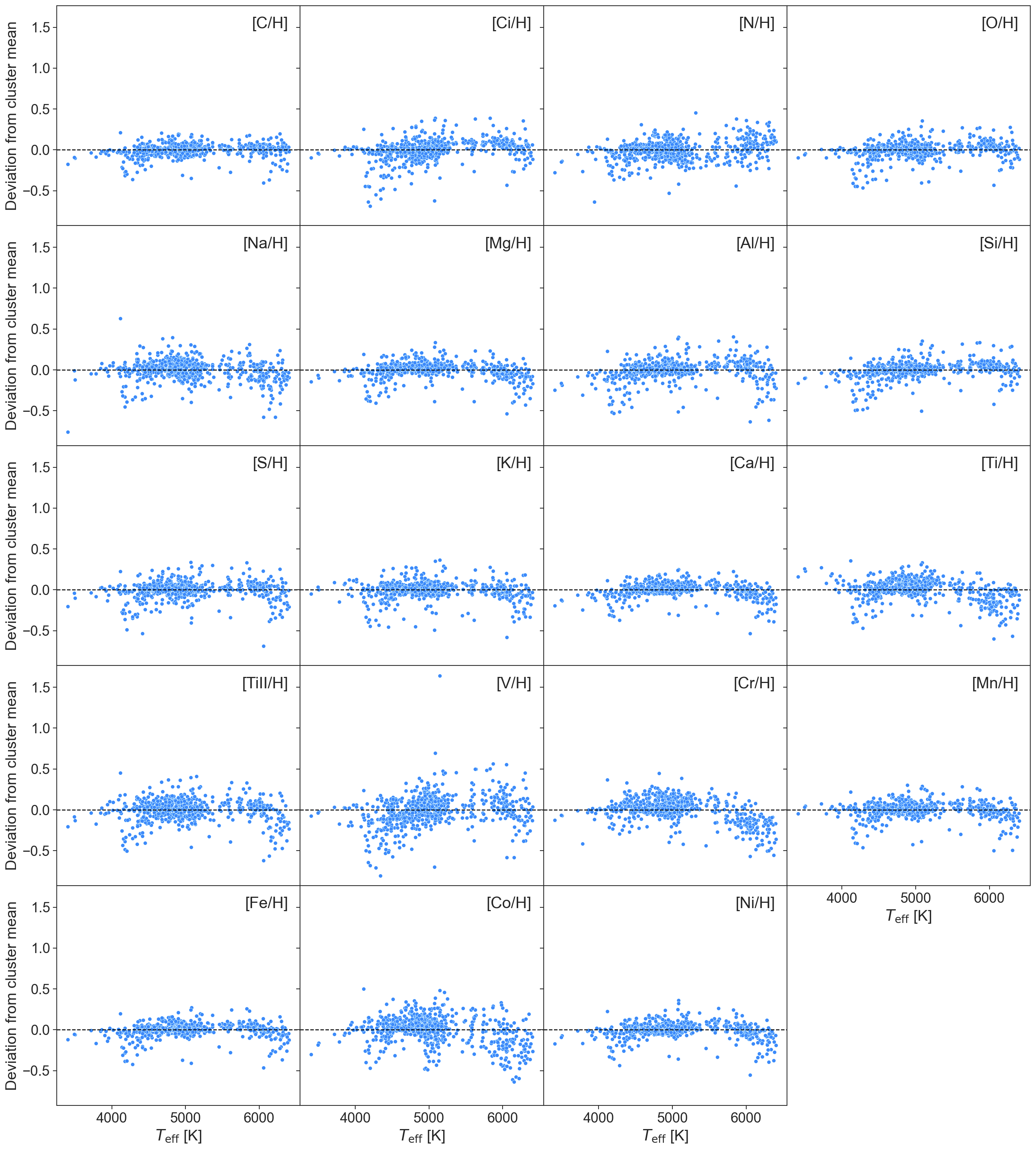} 
    \caption{Deviations from cluster mean abundance as a function of effective temperature for all the considered chemical elements. Each subplot displays a single element, as indicated in the top right corner of every panel, with deviations calculated relative to the mean abundance within each individual cluster.}
    \label{fig:trendsTeff}
    \end{figure}
\begin{figure}[ht]
    \centering
    \includegraphics[width = \linewidth]{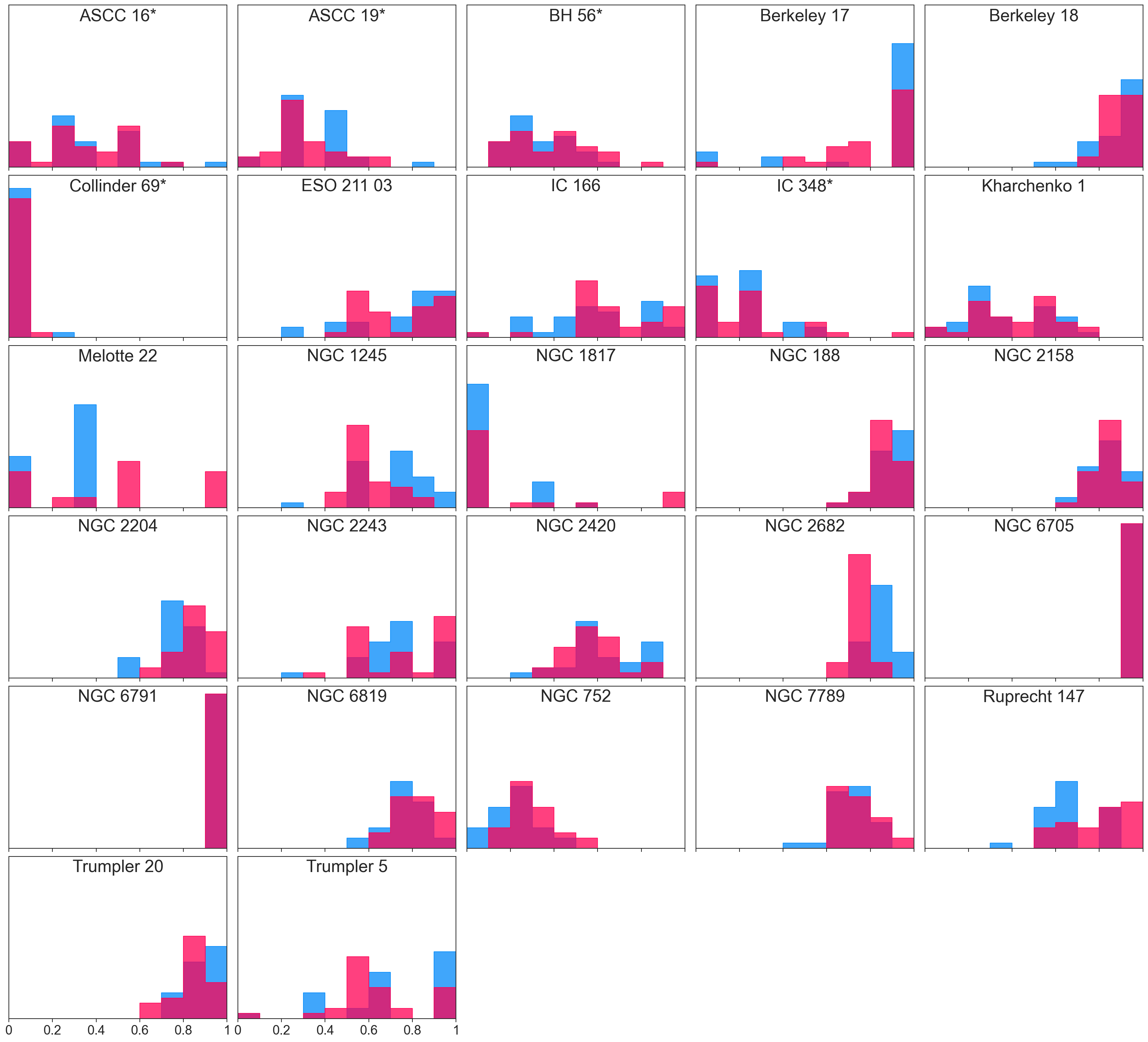}
    \caption{Distribution of recall (blue) and precision (red) for every high-quality cluster. We normalize each histogram to obtain an area equal to unity. (*) clusters not flagged as high-quality.}
    \label{fig:clusters_perf_distribution}
\end{figure}
\end{appendix}
\end{document}